# Spin-1/2 square-root operator equation with Coulomb potential – a perturbation analysis

Tobias Gleim

Seeking for a relativistic generalisation of the non-relativistic Schrödinger equation, one very soon arrives at equations with a square-root operator by having applied the quantum mechanical correspondence principle to the formula of relativistic energy. The problems of these equations are at least two fold: when coupled to an electromagnetic field, their relativistic invariance is not evident or even doubtful and due to their non-local character, it seems to be that they cannot be maintained mathematically in an easy way. For spin-1/2 particles, these difficulties can be overcome by the Dirac equation, which leads e.g. to the prediction of binding energies of an electron in a hydrogen atom that are compatible with experimental results up to the forth order of the fine structure constant, inclusively. Ignoring the problem with relativistic invariance, one may ask, if there exists a square-root equation, for which one can achieve the same good agreement with experiments for the latter physical system. It is going to be shown, that the answer to this question is affirmative and does not exceed the skills obtained in a course about non-relativistic quantum mechanics and physics of atoms, respectively.

An introductory course about relativistic quantum mechanics very often starts with the search for a correct relativistic wave equation for electrons, i.e. an equation that among other things reproduces the experimental results for the binding energies of an electron in a hydrogen atom. By means of the quantum mechanical correspondence principle and the formula of the relativistic energy, one soon arrives at the following eigenvalue equation for a free particle with mass $m$ and energy $E$ in position space representation (see e.g. [1, 2]):

$$E\psi(\vec{x}) = \pm\left(\sqrt{m^2 + \hat{\vec{p}}^2}\,\psi\right)(\vec{x}), \qquad (1)$$

wherein $\hat{\vec{p}} = -i\vec{\nabla}$ denotes the position space representation of the momentum vector. For spin-0 particles, the wave function $\psi$ is an ordinary scalar, but for spin-1/2 particles, it must be a two-spinor. The energy operator in (1) is apparently of non-local nature. The minimal coupling of (1) for a spin-0 particle with charge $e$ to an electromagnetic field with a scalar potential $V$ and a vector potential $\vec{A}$ (making now and in the following use of natural units, i.e. $\hbar = c = 1$),

$$E\psi(\vec{x}) = \pm\left(\sqrt{m^2 + (\hat{\vec{p}} - e\vec{A})^2}\,\psi\right)(\vec{x}) + V\psi(\vec{x}), \qquad (2)$$

shows that the relativistic invariance of the wave equation is rather doubtful or at least not evident (cf. [1]). Despite of these problems, (1) and (2), respectively, are of interest, because on the one hand even for spin-0 particles (1) possesses a positive definite probability density [3], which is no more the case, when (1) has been iterated to circumvent the difficulties with it and thus has become the well-known Klein-Gordon equation (see e.g. [1, 2]). On the other hand, (2) with a Coulomb potential (and $\vec{A} = 0$) appears as an approximation of the so called "Bethe-Salpeter equation" for spinless particles, with which bound states of e.g. fermion-antifermion pairs can be described within the framework of relativistic quantum field theory [4]. Thus, among other things, a perturbation analysis of the eigenvalue equation (2) with a Coulomb potential,

$$V(r) = -\frac{\alpha}{r}, \qquad (3)$$

with a small fine structure constant like $\alpha = e^2 \approx 1/137$ for the hydrogen atom, $\vec{A}=0$ and a scalar wave function has already been performed [5]: if one is only interested in terms with small powers of the fine structure constant $\alpha$, e.g. up to the order of $\alpha^4$ inclusively, one can start with a power series expansion of the square-root operator in (2),

$$\sqrt{1+\hat{x}} = 1 + \frac{1}{2}\hat{x} - \frac{1}{8}\hat{x}^2 + \ldots \qquad (4)$$

with $\hat{x} = \hat{p}^2/m$, and use ordinary perturbation theory of eigenvalues – although for still higher orders in $\alpha$, this is no longer possible [5].

Here, we want to restrict ourselves to an expansion of the square-root operator up to the order of $\alpha^4$, inclusively. But this time, we are not interested in the coupling of (1) for spin-0 particles to an electromagnetic field, but rather in the corresponding equation for spin-1/2 particles. First we have to make a guess for the coupling to an electromagnetic field, because (2) for a two-spinor $\psi$ would never reproduce Pauli's eigenvalue equation,

$$E\psi(\vec{x}) = \left( \frac{[\vec{\sigma} \cdot (\hat{\vec{p}} - e\vec{A})]^2}{2m} + V \right) \psi(\vec{x}), \qquad (5)$$

if one expands the square-root in (2) with the help of (4). Since Pauli's matrices $\vec{\sigma}$ appear in (5), one has to introduce them into (2) as well. Doing this, yields the following Hamiltonian:

$$\hat{H} = \sqrt{m^2 + (\vec{\sigma} \cdot (\hat{\vec{p}} - e\vec{A}(\vec{x})))^2} + V = \sqrt{m^2 + (\hat{\vec{p}} - e\vec{A})^2 - e\vec{\sigma} \cdot \vec{B}} + V \qquad (6\,a)$$

in a Schrödinger-like eigenvalue equation of the form

$$E\psi(\vec{x}) = \hat{H}\psi(\vec{x}), \qquad (6\,b)$$

where we have used

$$\sigma_k \sigma_j = \delta_{kj} + i\sum_{l=1}^{3}\varepsilon_{kjl}\sigma_l \qquad (7)$$

and

$$\vec{B} = \vec{\nabla} \times \vec{A}. \qquad (8)$$

But (6 a) does not help either, because (5) cannot be used to find the binding energies of an electron in a hydrogen atom, if they are to be valid up to the order of $\alpha^4$, inclusively: in (5), one has to set $\vec{A}$ to zero, which only reproduces the non-relativistic Schrödinger equation.

Reformulating Maxwell's equations by means of Pauli matrices, one gets

$$\left(\mathbf{1}\hat{p}_0 \mp \vec{\sigma} \cdot \hat{\vec{p}}\right)\Phi_\pm = \pm\left(\rho\mathbf{1} \pm \vec{\sigma} \cdot \vec{J}\right) \qquad (9)$$

with $\Phi_\pm = \vec{\sigma} \cdot (\vec{B} \mp i\vec{E})$ (containing besides the magnetic induction (8) also the electric field $\vec{E}$), a density of charge $\rho$ as well as a current $\vec{J}$ and an operator $\hat{p}_0 = i\partial_t$, as one can convince oneself by using (7). Therefore, (9) suggests replacing $\vec{\sigma} \cdot \vec{B}$ in (6) by $\Phi_\pm = \vec{\sigma} \cdot (\vec{B} \mp i\vec{E})$ in the spirit of a kind of „minimal" coupling with respect to the Maxwell equations in their formulation with Pauli matrices:

$$\hat{H} = \sqrt{m^2 + \left(\hat{\vec{p}} - e\vec{A}\right)^2 - e\vec{\sigma} \cdot \left(\vec{B} \mp i\vec{E}\right)} + V . \tag{10}$$

In (10), we have made a choice in favour of the positive sign of the square-root, because we want to address to particles. The other sign should be valid for anti-particles (or vice versa), in analogy to the spin-0 case. But the sign of $\vec{E}$ in (10) is still arbitrary: we do not know which one to choose, i.e. which of both is the "correct" one.

Hamiltonian (10) is going to be the starting point of our investigation. What we do next is to restrict ourselves to a Coulomb potential (3) and to set $\vec{A}$ to zero in (10):

$$\hat{H} = \sqrt{m^2 + \hat{\vec{p}}^2 \pm e\vec{\sigma} \cdot i\vec{E}} + V . \tag{11}$$

Expanding the square root in (11) with the aid of (4), we only take those terms into account that will come out to be of order $\alpha^4$ or lower, when performing a perturbation analysis for eigenvalues with the Hamiltonian obtained in that way. Finally, we can interpret that energy operator as a composition of the non-relativistic Schrödinger Hamiltonian plus a relativistic correction. Using this in a perturbation calculation, we would like to show, that the binding energies of an electron in a hydrogen atom can be reproduced up to the order of $\alpha^4$, inclusively.

*Derivation of a perturbation Hamiltonian*

Now we set $\hat{x}$ in (4) to the term $\left(\hat{\vec{p}}^2 \pm e\vec{\sigma} \cdot i\vec{E}\right)/m^2$ in order to obtain a formal expansion of the square-root operator in (11) and interpret $\hat{x}^2$ as two subsequent applications of the operator $\hat{x}$. Calculated in this way, $\hat{x}^2$ leads to contributions in (4) that among other things contain terms in $\vec{E}^2$ or terms with two operators $\hat{\vec{p}}$ and the electric field $\vec{E}$. Since first order perturbation theory for eigenvalues is based on taking the expectation value of a Hamiltonian by means of the states of the unperturbed system, it is useful to realise that expectation values of powers of the radius *r* being calculated with the help of the non-relativistic energy eigenfunctions of the hydrogen atom show the following proportionality relation with respect to the fine structure constant $\alpha$ (see e.g. [7]):

$$\left\langle r^k \right\rangle \propto \alpha^{-k} \tag{12}.$$

Therefore, we can consider $\hat{\vec{p}}$ to be of order $\alpha$ as well as $e\vec{E}$ to be of order $\alpha^3$, the latter because of

$$e\vec{E} = -\vec{\nabla}V = -\alpha \frac{\vec{x}}{r^3} \tag{13}.$$

For we only want to take into account terms up to the order of $\alpha^4$ inclusively, we must neglect terms in $\vec{E}^2$ or two operators $\hat{\vec{p}}$ together with $\vec{E}$. In the spirit of the perturbation analysis that we are striving for, this yields the following approximation of the Hamiltonian (11) being valid up to the order of $\alpha^4$, inclusively:

$$\hat{H} = \frac{\hat{\vec{p}}^2}{2m} + V(r) \pm i\frac{e}{2m}\vec{\sigma} \cdot \vec{E} - \frac{1}{2m}\left(\frac{\hat{\vec{p}}^2}{2m}\right)^2 , \tag{14}$$

where we have absorbed the term with the mass *m* into the definition of the Hamiltonian $\hat{H}$ on the left hand side.

Introducing the angular momentum operator

$$\hat{\vec{L}} = \vec{x} \times \hat{\vec{p}},\qquad(15)$$

with which one is able to derive the useful equation

$$\hat{\vec{p}}^2 = -\frac{1}{r}\partial_r^2 r + \frac{\hat{\vec{L}}^2}{r^2},\qquad(16)$$

(14) can be represented in the subsequent form:

$$\hat{H} = -\frac{1}{2m}\frac{1}{r}\partial_r^2 r - \frac{\alpha}{r} + \frac{1}{2m}\frac{\hat{\vec{L}}^2 \mp i\alpha\vec{\sigma}\cdot\hat{e}_r - \alpha^2}{r^2} - \frac{1}{2m}\left(\frac{\hat{\vec{p}}^2}{2m}\right)^2 + \frac{1}{2m}\frac{\alpha^2}{r^2},\qquad(17)$$

where we have used the abbreviation $\hat{e}_r = \vec{x}/r$ and subtracted $\alpha^2/(2m r^2)$ from the term with $\hat{\vec{L}}^2$ in order to add it again at the end of (17). The latter term is admissible, because it is only of order $\alpha^4$. In order to be able to work with Hamiltonian (17), we have to use a special sort of angular momentum algebra that can be found in [6]. In the subsequent section, we are going to list those results from it that will be used later.

*Angular momentum algebra*
Defining an operator

$$\hat{\Lambda} = -\left(\vec{\sigma}\cdot\hat{\vec{L}} + 1\right) \mp i\alpha\vec{\sigma}\cdot\hat{e}_r,\qquad(18)$$

one obtains the equation

$$\hat{\Lambda}^2 = \left(\vec{\sigma}\cdot\hat{\vec{L}} + 1\right)^2 - \alpha^2\qquad(19)$$

because of the identity

$$\left(\vec{\sigma}\cdot\hat{\vec{L}} + 1\right)\vec{\sigma}\cdot\hat{e}_r + \vec{\sigma}\cdot\hat{e}_r\left(\vec{\sigma}\cdot\hat{\vec{L}} + 1\right) = 0.\qquad(20)$$

By means of (7) and (15), the validity of (20) as well as of the following equation can be shown:

$$\vec{\sigma}\cdot\hat{\vec{L}}\left(\vec{\sigma}\cdot\hat{\vec{L}} + 1\right) = \hat{\vec{L}}^2.\qquad(21)$$

With the aid of (18), (19) and (21), it is rather easy to prove the subsequent identity:

$$\hat{\Lambda}\left(\hat{\Lambda} + 1\right) = \hat{\vec{L}}^2 \mp i\alpha\vec{\sigma}\cdot\hat{e}_r - \alpha^2\qquad(22)$$

which is the numerator of the third term in the Hamiltonian (17). If one now introduces a total angular momentum operator

$$\hat{\vec{J}} = \hat{\vec{L}} + \frac{1}{2}\vec{\sigma},\qquad(23)$$

one can find eigenfunctions $|l\tfrac{1}{2}jm\rangle$ of its square and of its 3-component (with eigenvalues $j(j+1)$ and $m$, respectively) that are simultaneous eigenfunctions of the squared orbital angular momentum operator $\hat{L}^2$ (with eigenvalue $l(l+1)$) and of the square of the spin operator $\tfrac{1}{2}\vec{\sigma}$ (with eigenvalue ¼), too. With the help of a Clebsch-Gordan expansion, those wave functions can be constructed in a basis $|l\tfrac{1}{2}m_l m_s\rangle$ of the eigenfunctions of the square of the orbital and of the spin angular momentum operators as well as their 3-components (whose eigenvalues are denoted by $m_l$ and $m_s$, respectively). For our perturbation analysis, we regard the latter wave functions essentially as a linear combination of angular and spin parts of the solutions of the non-relativistic Schrödinger equation for the hydrogen atom.

With a bit of angular momentum algebra, the application of the operator $\left(\vec{\sigma}\cdot\hat{\vec{L}}+1\right)$ to the wave functions $|l\tfrac{1}{2}jm\rangle$ now yields

$$\left(\vec{\sigma}\cdot\hat{\vec{L}}+1\right)|l\tfrac{1}{2}jm\rangle = \pm(j+\tfrac{1}{2})|l\tfrac{1}{2}jm\rangle \qquad (24)$$

with $j = l \pm \tfrac{1}{2}$. This equation is a result of the coupling of the orbital angular momentum vector to the spin, i.e.

$$\hat{\vec{J}}^2 - \hat{\vec{L}}^2 - \left(\tfrac{1}{2}\vec{\sigma}\right)^2 = 2\hat{\vec{L}}\cdot\tfrac{1}{2}\vec{\sigma} = 2\hat{L}_3\cdot\tfrac{1}{2}\sigma_3 + \hat{L}_+\cdot\tfrac{1}{2}\sigma_- + \hat{L}_-\cdot\tfrac{1}{2}\sigma_+, \qquad (25)$$

where we have used (23) and introduced raising (with index +) and lowering (with index -) operators, i.e. a linear combination of 1- and 2- components of $\hat{\vec{L}}$ and $\vec{\sigma}$ of the kind

$$\hat{L}_\pm = \hat{L}_1 \pm i\hat{L}_2,\ \sigma_\pm = \sigma_1 \pm i\sigma_2, \qquad (26)$$

as well as the 3-components (with index 3) of the orbital and spin angular momentum operators. The algebra that those lowering and raising operators obey is presented in many textbooks about quantum mechanics or physics of atoms (see e.g. [1, 6, 7]). Its application leads to (24).
From (19) and (24), we can conclude the following eigenvalue equations:

$$\hat{\Lambda}^2 |l\tfrac{1}{2}jm\rangle = \left[(j+\tfrac{1}{2})^2 - \alpha^2\right]|l\tfrac{1}{2}jm\rangle, \qquad (27)$$

$$\hat{\Lambda}|l\tfrac{1}{2}jm\rangle = \mp\sqrt{(j+\tfrac{1}{2})^2 - \alpha^2}\,|l\tfrac{1}{2}jm\rangle \qquad (28)$$

with $j = l \pm \tfrac{1}{2}$. The signs in (28) follow from the limit $\alpha \to 0$ together with (18) and (24). Denoting the eigenvalue of $\hat{\Lambda}(\hat{\Lambda}+1)$ by $\lambda(\lambda+1)$, one can extract $\lambda$ form (27) and (28) as

$$\lambda = l - \varepsilon_j, \qquad (29\text{ a})$$

$$\varepsilon_j = j + \tfrac{1}{2} - \sqrt{(j+\tfrac{1}{2})^2 - \alpha^2}. \qquad (29\text{ b})$$

Now, we are able to perform calculations with Hamiltonian (17) in the framework of our perturbation analysis for energy eigenvalues of an electron bound in a hydrogen atom, what we are going to show in the next section.

*Perturbation analysis for eigenvalues*

As already mentioned, for our perturbation analysis for eigenvalues, we are going to choose the solutions of the non-relativistic Schrödinger equation for the hydrogen atom as wave functions, but with a linear combination of their angular and spin parts, that we have denoted by $\left|l\tfrac{1}{2}jm\right\rangle$ in the last section. We will call the resulting wave functions $\psi_{nl\frac{1}{2}jm}$. When Hamiltonian (17) acts on these wave functions, the operator (22) – that is a part of (17) – acts on $\left|l\tfrac{1}{2}jm\right\rangle$ and so we can use (27), (28) and (29). But the term of (17), in which (22) appears, already contains a factor $1/r^2$ that will turn out to be of order $\alpha^2$ (cf. (12)). Since we are only interested in terms up to the order of $\alpha^4$ inclusively, we have to expand the square-root of (29 b) in $\alpha$ and thereof only retain powers up to the order of $\alpha^2$ inclusively:

$$\varepsilon_j \approx \frac{\alpha^2}{2j+1} \tag{30}$$

Hence the eigenvalues of (22) are approximately equal to

$$\lambda(\lambda+1) \approx l(l+1) - \alpha^2 \frac{l+\tfrac{1}{2}}{j+\tfrac{1}{2}}, \tag{31}$$

if all terms of higher order than $\alpha^2$ are neglected. In this approximation, Hamiltonian (17) acting on the wave functions $\psi_{nl\frac{1}{2}jm}$ yields the subsequent equation:

$$\hat{H}\psi_{nl\frac{1}{2}jm} \approx \left[\frac{1}{2m}\left(-\frac{1}{r}\partial_r^2 r + \frac{l(l+1)}{r^2}\right) + V(r) - \frac{1}{2m}\left(\frac{\hat{p}^2}{2m}\right)^2 - \frac{1}{2m}\alpha^2 \frac{l+\tfrac{1}{2}}{j+\tfrac{1}{2}}\frac{1}{r^2} + \frac{1}{2m}\frac{\alpha^2}{r^2}\right]\psi_{nl\frac{1}{2}jm}. \tag{32}$$

(32) contains a Hamiltonian that is an approximation of (11) being valid up to the order of $\alpha^4$ inclusively. The first three terms in (32) belong to the non-relativistic Hamiltonian of the unperturbed system:

$$\hat{H}_0 = \frac{\hat{p}^2}{2m} + V(r), \tag{33}$$

whereas the remaining terms represent relativistic corrections to it and can be combined to a perturbation Hamiltonian:

$$\hat{H}_1 = -\frac{1}{2m}\left(\frac{\hat{p}^2}{2m}\right)^2 - \frac{1}{2m}\alpha^2 \frac{l+\tfrac{1}{2}}{j+\tfrac{1}{2}}\frac{1}{r^2} + \frac{1}{2m}\frac{\alpha^2}{r^2}. \tag{34}$$

If we now use first order perturbation theory for eigenvalues, we have to calculate the following expectation values of the sum of (33), delivering the non-relativistic energies $E_n$ of an electron in a Coulomb potential, and (34):

$$E_{nl} \approx \left\langle \psi_{nl\frac{1}{2}jm}\left|\hat{H}_0 + \hat{H}_1\right|\psi_{nl\frac{1}{2}jm}\right\rangle = E_n + \Delta E_1, \tag{35}$$

with

$$\Delta E_1 = \left\langle \psi_{nl\frac{1}{2}jm} \middle| \hat{H}_1 \middle| \psi_{nl\frac{1}{2}jm} \right\rangle = \left\langle \hat{H}_1 \right\rangle. \tag{36}$$

The first term in (34) can be expressed by means of (33) whose eigenvalues are known to be

$$E_n = -\frac{m\alpha^2}{2n^2}. \tag{37}$$

The contribution of this term to (36) amounts to (see e.g. [7])

$$\Delta E_{10} = -\frac{1}{2m}\left[E_n^2 + 2E_n\langle V(r)\rangle + \langle V^2(r)\rangle\right] = -E_n \frac{\alpha^2}{n^2}\frac{3}{4} + E_n \frac{\alpha^2}{n}\frac{1}{(l+1/2)}, \tag{38}$$

where we have used

$$\left\langle \frac{1}{r} \right\rangle_{nlm} = \frac{1}{a_0 n^2} \tag{39}$$

and

$$\left\langle \frac{1}{r^2} \right\rangle_{nlm} = \frac{1}{a_0^2 n^3 (l+1/2)} \tag{40}$$

with the Bohr radius $a_0 = 1/(m\alpha)$. For the last two terms in (34) and their expectation values in (36), we only have to apply (40), what results into

$$\Delta E_{11} = E_n \frac{\alpha^2}{n}\frac{1}{j+\frac{1}{2}} - E_n \frac{\alpha^2}{n}\frac{1}{(l+1/2)}. \tag{41}$$

In $\Delta E_1 = \Delta E_{10} + \Delta E_{11}$ the last term of (41) and the last term of (38) cancel out each other, thus the energy eigenvalues (35) become

$$E_{nj} \approx E_n\left[1 - \frac{\alpha^2}{n^2}\left(\frac{3}{4} - \frac{n}{j+1/2}\right)\right] \tag{42}$$

that is valid up to the order of $\alpha^4$, inclusively.

*Discussion and Conclusions*

We have shown that up to the order of $\alpha^4$ inclusively, Hamiltonian (11) for an electron in a Coulomb potential (and containing a square-root operator) can be approximated by (14) and the Hamiltonian in (32), respectively. The latter can be used in an eigenvalue analysis that, within the framework of this approximation, yields the correct binding energies (42) of an electron in a Coulomb potential. Using the well-known Dirac equation, one gets energy eigenvalues

$$E_{nj} = \frac{m}{\sqrt{1 + \frac{\alpha^2}{(n-\varepsilon_j)^2}}}. \tag{43}$$

Expanding this up to the same order of the fine structure constant, one obtains (42) again (see e.g. [1, 2]). If one performs an approximation of the Dirac-Hamiltonian, the subsequent Hamiltonian is obtained (see e.g. [1, 2]):

$$\hat{H} = \frac{1}{2m}\hat{\vec{p}}^2 + V - \frac{1}{2m}\left(\frac{\hat{\vec{p}}^2}{2m}\right)^2 - \frac{1}{8m^2}\hat{\vec{p}}^2 V + \frac{1}{4m^2}\vec{\sigma}\cdot\vec{\nabla}V\times\hat{\vec{p}} \tag{44}$$

which can be used in a perturbation calculation for eigenvalues, too, and leads to (42) again. (42) is the formula of the binding energies for an electron in a hydrogen atom that reproduces the correct experimental results – at least apart from contributions that can only be described by means of quantum electrodynamics. Thus with respect to the hydrogen atom, there seem to be no experimental objections against Hamiltonian (11) and the way it couples to an electromagnetic field that is different from the "normal" minimal coupling scheme, because electromagnetic field terms associate with the mass term under the square-root (cf. also [12] that deals with square-root equations for four-spinors). But of course, the already mentioned theoretical objections against (11) are still valid and it is even doubtful, whether (10) or (11) can be Hermitian operators at all due to the term $i\vec{\sigma}$ therein, which is not Hermitian either. On the other hand, the sign of the term $\pm i\vec{\sigma}\cdot e\vec{E}$ within those Hamiltonians does not play any role with respect to (42), because we started by taking into account both signs and ended with the same energies (42) for each sign. Therefore, our analysis was not able to show under which circumstances one sign has to be preferred with respect to the other one: for a particle (and also its antiparticle), two ways of coupling to a Coulomb potential are possible.

It is astonishing, that the term $\pm i\frac{e}{2m}\vec{\sigma}\cdot\vec{E}$ in (14) leads to the correct spin-orbit coupling – described by the last term in (44) – as well as comprises the effects of the Darwin term in (44), which takes into account the so called trembling motion of the electron, usually being interpreted as an effect of an interference of particle and anti-particle contributions. The latter is rather surprising, because with (11), we started with a Hamiltonian being only valid for particles, thus a mixture of particle and anti-particle contributions seems to be impossible.

Maybe, the results (42) are less surprising, if we realise that the exact binding energies (43) of an electron in a Coulomb potential can also be deduced from the subsequent Klein-Gordon-like eigenvalue equation for 2-spinors $\psi_\pm$:

$$(E-V)^2\psi_\pm = \left(m^2 + \hat{\vec{p}}^2 \pm i\vec{\sigma}\cdot e\vec{E}\right)\psi_\pm, \tag{45}$$

because (45) is a direct consequence of the Dirac equation, what is shown in [6]. But the iteration of our Schrödinger-like equation with Hamiltonian (11) does not give (45), because the momentum operator does not commute with the Coulomb potential or the electric field. On the other hand, of course, one would immediately guess, that Hamiltonian (14) is a non-relativistic limit of (45) with a (small) relativistic correction of the kinetic energy operator. Actually, this is not exactly the case, but one can show, that the extra terms that arise with respect to (14) cancel out each other when calculating the expectation value (35) – the straightforward but rather lengthy proof of this should be given elsewhere. In [6], it is elucidated that the wave function $\psi_+$ belongs to particles and $\psi_-$ to anti-particles (or vice versa) and that the energy eigenvalues turn out to be the same, i.e. are independent of the sign in the term $\pm i\vec{\sigma}\cdot e\vec{E}$ of (45). Therefore, (45) suffers from the same "arbitrariness of sign" problem as (14). In (45), this can be understood well, because it has been derived by starting with an equation for a 4-spinor, namely the Dirac equation, and ends with two 2-spinor equations, i.e. (45). The other way round, one could also say that the sign problem of (45) can be solved by the introduction of 4-spinors, i.e. by means of the Dirac equation. In this context, it is rather amusing to see that even the Maxwell equations in the form of (9) have two different versions with respect to the signs in the differential operator $\left(\mathbf{1}\hat{p}_0 \mp \vec{\sigma}\cdot\hat{\vec{p}}\right)$ and in the 4-current $\rho\mathbf{1}\pm\vec{\sigma}\cdot\vec{J}$, but which are entirely equivalent. Hence one could conceive a similar method to get rid of this "arbitrariness of sign" problem there, too: i.e. using Dirac's $4\times 4$ gamma- rather than Pauli's $2\times 2$-matrices.

Since (45) is of the same type as the Klein-Gordon equation, it has the same benefits, but also suffers from the same problems as the latter one: it is obviously relativistic invariant – even when coupled to an electromagnetic field (at least when also taking a Vector potential and the magnetic field, respectively, into account) – but its probability density is not positive definite either; it rather describes two particles at the same time, namely particles and anti-particles, than that it gives a one-particle interpretation.

The Dirac equation seems to solve all problems but the last one: the postulate of a "Dirac sea" must be introduced, to make a one-particle interpretation possible: see e.g. [1, 2, 8]. It is often stated, that one gets rid of this within the framework of quantum field theory. But this is actually not true – instead, one introduces the same postulate into quantum field theory again: see. e.g. [9, 10].

For the square-root equations, because of their Schrödinger-like form, positive definiteness of the probability density seems to be no problem. Even relativistic invariance is unproblematic, as long as (or maybe even *only when*) free particles are described: see e.g. [3, 11]. A Dirac sea postulate for spin-1/2 particles is not necessary, because there is a separate equation for both particles and anti-particles.

What comes out, is that many equations can lead to the correct predictions of experimental results – as it is here the case for the binding energies of an electron in a hydrogen atom. In the end, they do not seem to be so different any more, but at least with the here already mentioned experimental results for the hydrogen atom, one cannot decide, which is the "correct" one. Instead, one needs theoretical arguments in order to make a decision: e.g. Hermiticity of the energy operator to obtain always real eigenvalues (independent of the system regarded), positive definiteness of the probability density and a one-particle description in order to make a probability interpretation possible, relativistic invariance of both equations: the one for a free particle and the one for a particle in an electromagnetic field etc. But unfortunately, neither of the equations regarded have been able to fulfil all the theoretical expectations – at least not without posing additional postulates.

Therefore, it is not so inept to consider square-root operator equations again, even if they mean a substantial mathematical burden, which was presumably one of the main reasons in the past to discard them: even in our simple eigenvalue analysis, the approximations that were performed need a more thorough mathematical investigation.

Within this paper, more questions emerged than could be answered: intrinsically connected with our analysis seems to be a further question, namely, whether a 4- or a 2-spinor should be used to describe electrons (or more general: fermions) as well as whether Dirac's $4\times 4$- rather than Pauli's $2\times 2$-matrices should be used for the description of photons: the former ones could help to get rid of an arbitrary choice of sign within the corresponding wave equations. But it is not clear, if this is really a physical or just an aesthetic aspect.